\documentclass[aps,groupedaddress,twocolumn,prb,showpacs]{revtex4}

\draft

\usepackage{graphicx}
\usepackage{dcolumn}
\usepackage{amsmath}
\usepackage{bm}
\begin{document}

\title {Long-range radiative interaction between semiconductor quantum dots}
\author{G. Parascandolo, V. Savona}
\affiliation{Institute of Theoretical Physics, Ecole Polytechnique F\'ed\'erale de Lausanne (EPFL), CH-1015 Lausanne, Switzerland}

\date{\today}

%%%%%%%%%%%%%%%%%%%%%%%%%%%%%%%%%%%%%%%%%%%%%%%%%%%%%%%%%%%%%%

\begin{abstract}
We develop a Maxwell--Schr\"{o}dinger formalism in order to describe the radiative interaction mechanism between semiconductor quantum dots. We solve the Maxwell equations for the electromagnetic field coupled to the polarization field of a quantum dot ensemble through a linear non--local susceptibility and compute the polariton resonances of the system. The radiative coupling, mediated by both radiative and surface photon modes, causes the emergence of collective modes whose lifetimes are longer or shorter compared to the ones of non--interacting dots. The magnitude of the coupling and the collective mode energies depend on the detuning and on the mutual quantum dot distance. The spatial range of this coupling mechanism is of the order of the wavelength. This coupling should therefore be accounted for when considering quantum dots as building blocks of integrated systems for quantum information processing.
\end{abstract}

\maketitle

\section{Introduction}

Exciton--polaritons are the basic optical excitations of any semiconductor system. The interband optical polarization of a semiconductor is never an isolated degree of freedom. Rather, the exciton states are always coupled to the electromagnetic field via the linear radiation--matter interaction Hamiltonian. Exciton--polaritons are the resulting eigenmodes of the Maxwell equations coupled to the material equations describing the excitons in a semiconductor structure. The polariton concept in bulk semiconductors was first introduced by Hopfield \cite{Hopfield_PR_1958}. Bulk exciton--polaritons are mixed modes of one exciton and one photon mode having the same momentum, as imposed by translational invariance. This one--to--one selection rule results into a strong mixing and an energy--dispersion displaying the anticrossing typical of normal--mode coupling \cite{Andreani_Review}. In GaAs, the normal--mode splitting at resonance is 16 meV, larger than the exciton binding energy.

In systems with reduced dimensionality, such as quantum wells and quantum wires, the partial breaking of the translational symmetry allows coupling of excitons to a continuum of photon modes. The polariton picture is consequently modified and a polariton becomes a resonance of a discrete exciton state linearly coupled to a photon continuum, analogously to a Fano resonance \cite{Fano_PR_1961}. In this case the importance of the coupling, expressed by the magnitude of the polariton self--energy correction to the bare exciton energy, is considerably smaller. In quantum wells \cite{Andreani_Review,Tassone_NCimento_1990,Citrin_PRB_1993,Jorda_PRB_1993,Jorda_PRB_1994}, the polariton resonance implies a finite exciton radiative lifetime which is of the order of 10 ps in typical GaAs quantum wells, and a negligible shift of the exciton energy. A similar effect is predicted in quantum wires \cite{Citrin_PRL_1992,Citrin_PRB_1993_2,Tassone_PRB_1995}.

When the dimensionality of the electromagnetic field is also reduced, e.g. in semiconductor planar microcavities, the one--to--one coupling typical of a bulk semiconductor is recovered and strongly--coupled polaritons with full exciton--photon mixing characterize the optical spectrum \cite{Weisbuch_PRL_1992,Houdre_PRL_1994,Savona_SSC_1995,Savona_PRB_1996}.

The question naturally arises, whether the polariton concept is of some relevance in the case of quantum dots (QDs), where the electron--hole system is fully confined in the three spatial dimensions. In this case we might distinguish between two effects of the radiation--matter coupling. The first is the self--energy of a single QD coupled to the electromagnetic field, resulting in a finite radiative lifetime and an energy shift of the QD levels. This effect is well established and several theoretical estimates of the radiative lifetime of QDs have been proposed in the past years \cite{Bockelmann_PRB_1993,Citrin,Kavokin_APL_2002}. The second is the radiative coupling between different QDs. This process might be depicted as a multiple emission and reabsorption of a photon by QDs in a many--QD system, eventually giving rise to collective modes of several QDs. This picture implies that the exciton state of a single QD is no longer an eigenstate and excitation of one exciton in a QD would result in a transfer of excitation to other QDs, similar to a system of coupled harmonic oscillators. A similar effect has already been suggested and theoretically characterized in the case of quantum well excitons localized by interface disorder \cite{Zimmermann,Savona}. 

The excitation transfer mechanisms between polarizable systems such as molecules or semiconductor QDs are usually grouped into two categories, depending on whether they occur via overlapping wave functions of the spatially separated systems or via long--range interactions. The propotype of these latter case is the electrostatic dipole--dipole interaction, frequently referred to as {\em F\"orster energy transfer} \cite{Foerster}. 
As an example, the F\"orster rate for the excitation transfer between two QDs has been estimated \cite{Govorov03} in the range of $10^{-2}$ to $10^{-3}\,\mbox{ps}^{-1}$ for InP QDs with interdot distance of 7 nm. This mechanism has been experimentally characterized in the case of closely spaced QD systems \cite{Kagan96}. The important feature of the F\"orster mechanism, however, is the dependence of the transfer rate on the distance. Being a dipole--dipole type interaction, its rate within Fermi golden rule is proportional to the fourth power of the dipole moment matrix element and decays as the sixth power of the distance \cite{Govorov03}. On a more general ground, in addition to the original F\"orster scheme, all excitation transfer mechanisms based on electromagnetic interaction are expected to be of some importance. The polariton mechanism that we address here involves an excitation transfer mediated by the transverse electromagnetic field. 
Within a perturbation picture, which considers only one emission--absorption process, this mechanism results in the transfer of electron--hole excitation between distant QDs mediated by the emission and reabsorption of a propagating photon. The coupling strength of this process is therefore expected to be small, if compared to other proposed coupling mechanisms which involve tunneling of the electron--hole wave function \cite{Bayer_Science_2001,Piermarocchi_PRL_2002,Langbein_PRL_2003}, or even to the F\"orster mechanism at short distance. This is basically due to the small absorption cross section of typical semiconductor QDs. However, the polariton coupling is also expected to have a long spatial range, of the order of a few photon wavelengths. Its rate turns out to be proportional to the square of the dipole moment matrix element and decays as the inverse of the distance, as the transfer is mediated by a propagating field in two dimensions. In addition, as for the F\"orster mechanism, it requires that the two energy levels involved in the excitation transfer have resonant spectra. Although, as it will turn out, typical transfer rates derived here are in the range of $10^{-3}$ to $10^{-4}\,\mbox{ps}^{-1}$, because of its spatial range the mechanism we study must be considered as complementary to the F\"orster coupling. In particular this long--range coupling might play an important role in the increasingly sought applications of QDs in quantum information processing \cite{Loss_PRB_1999,Imamoglu_PRL_1999,Quiroga99,Biolatti_PRL_2000,Ulloa_PRB_2004,Lovett_PRB_2003,Borri_PRB_2002}. In presence of radiative interaction, in fact, even very distant QDs cannot be considered as isolated systems.

In this work, in order to describe the radiative coupling mechanism, we develop a full Maxwell--Schr\"{o}dinger formalism for a system of many QDs coupled to the electromagnetic field. We model a QD having cylindrical shape and compute the ground electron--hole pair state within the effective mass scheme. We numerically solve the Maxwell equations for the electromagnetic field coupled to the polarization field of a QD ensemble, in order to compute the polariton resonances of the system. The linear susceptibility is obtained by means of the standard linear response theory. We initially address the case of two QDs and inspect how the coupling depends on the detuning and the mutual QD distance. Afterwards, we consider a system of many QDs randomly distributed on a plane and we determine the eigenenergies of the coupled equations. The collective modes of the system display modified radiative lifetimes, some of them being strongly {\em sub--radiant} or {\em super--radiant} with respect to the bare QDs lifetime. This is analogous to the case of polaritons in multiple quantum wells \cite{Citrin_PRB_1994,Andreani_PLA_1994}, although the QDs in our case are randomly distributed in space rather than ordered in a superlattice. We apply our model to the realistic cases of Stranski--Krastanov--grown InAs QDs and CdSe QDs resulting from interface fluctuations in narrow quantum wells. In the CdSe case the effect of radiative coupling is sizeable and a considerable number of collective modes have lifetimes about twice as long or short than the uncoupled case.

The article is organized as follows. In Sec. II, starting form the Maxwell equations and a linear non-local susceptibility, we analitically derive the eigenmode equations that holds in presence of radiative coupling. Sec. III contains the results of the numerical diagonalization of the problem in the cases of two-- and many--QD systems, followed by a discussion of the computed data. In Sec. IV we present some concluding remarks. Finally, Appendix A contains the details of the QD model used to derive the electron--hole pair wave functions that were used in this work, while Appendix B contains the detailed derivation of the radiative coupling tensor between two QDs.

\section{Theory}

The semiclassical model of QD interband excitation in interaction with the electromagnetic field is based on the solution of the Maxwell equations coupled to a nonlocal linear susceptibility which accounts for the interband optical transition. This is done in full analogy with the polariton formalism in bulk semiconductors and heterostructures~\cite{Andreani_Review,Tassone_NCimento_1990}. We restrict in what follows to the transition between the semiconductor ground state and the ground electron--hole pair state (that is the first excited state) in each QD. Within the effective mass approach, the linear susceptibility tensor (in what follows, tensors are indicated by a ``hat'') deriving from the linear response theory \cite{Kubo} is
\begin{equation} \label{Chi}
\hat{\bm{\chi}}\left({\bf r},{\bf r}^\prime,\omega\right)=
\frac{\mu_{cv}^2}{\hbar}\sum_\alpha
\frac{\Psi^{ }_\alpha\left({\bf r},{\bf r}\right)
\Psi^*_\alpha\left({\bf r^\prime},{\bf r^\prime}\right)}
{\omega_\alpha-\omega-i0^+}
\left(
\begin{array}{ccc}
1 & 0 & 0 \\
0 & 1 & 0 \\
0 & 0 & 0
\end{array}
\right)\,.
\end{equation}
The susceptibility is non--local in the three spatial coordinates, as expected from the breaking of traslational invariance. In Eq.~(\ref{Chi}), $\mu_{cv}$ is the dipole matrix element of the interband optical transition~\cite{Andreani_Review}. The quantities $\hbar\omega_\alpha$ and $\Psi_\alpha\left({\bf r}_e,{\bf r}_h\right)$ are respectively the electron--hole pair energy and wave function in the $\alpha$--th dot. We assume an electron--hole pair wave function which is factorized in its electron and hole parts, thus neglecting the electron--hole Coulomb correlation. It is reasonable to assume~\cite{Stier_PRB_1999,Zimmermann_ICPS} that for strongly confined systems the Coulomb correlation induces only a moderate quantitative change in the optical transition probability amplitude. Given the very simple description of the QD in this work, this quantitative effect can be accounted for by adjusting the interband matrix element in order to reproduce e.g. the single--QD radiative recombination rate. Note that in expression~(\ref{Chi}) the wave function $\Psi_\alpha$ is evaluated at ${\bf r}_e={\bf r}_h$, according to the effective mass theory of the interband optical transition \cite{Andreani_Review}. By introducing the susceptibility tensor (\ref{Chi}) in this particular matrix form,we are implicitely considering the electron--heavy--hole optical transition in a semiconductor with cubic lattice symmetry. In this case, in analogy with a quantum well~\cite{Savona}, only the $x$-- and $y$--components of the interband electron--hole polarization vector are coupled to the electromagnetic field, resulting in the particular shape of the susceptibility tensor~(\ref{Chi}). It is therefore possible to analytically solve the uncoupled z--component of the Maxwell equations and to effectively reduce to a two--dimensional problem. In this planar geometry there are two independent states of the interband polarization vector, that correspond to excitons with spins oriented along the x-- and y--direction, respectively. By using simple Lorentz resonances in (\ref{Chi}), we assume that the nonradiative lineshape of each QD is a Dirac delta function. Recently it was found that non--perturbative coupling of the exciton with acoustic phonons is responsible for a broad phonon--assisted contribution to the nonradiative QD lineshape \cite{Zimmermann_ICPS,Krummheuer02,Borri01}. However, at low temperatures the phonon--assisted part of the line tends to be small, especially for low quantum confinement. The zero--phonon part of the line on the other hand is only affected by the so called ``pure dephasing''. It has been recently shown that pure dephasing in QDs is almost exclusively due to the radiative recombination rate \cite{Langbein04} which is also an outcome of the present approach. We will restrict to a simple Lorentz lineshape for the uncoupled QD and assume that our results apply to the zero--phonon part of the interband excitation.

\begin{figure}[h]
\begin{center}
\includegraphics[width=8cm]{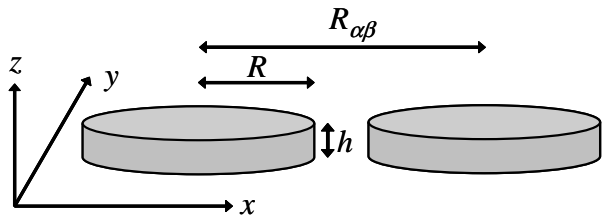}
\caption{Schematic diagram of the cylindrical dot ensemble. h is the height of a QD in the $z$--direction,  R its radius in the $(x,y)$--plane,  and 
$R_{\alpha\beta}$ is the distance between the centers of the two QDs.}
\label{two_dots}
\end{center}
\end{figure}

The QD we are considering has cylindrical shape with radius $R$ and height $h$ and is assumed to have a small aspect ratio $h/R$, as occurring for most real QD systems~\cite{Piermarocchi_Science_2003,Bonadeo_PRL_1998,Grundmann_PRB_1995,Hartmann_PRL_2000,Baier_APL_2004}. We are therefore treating a quasi two--dimensional system with the QD lying on the $(x,y)$--plane, as illustrated in Fig.~\ref{two_dots} in the case of two QDs labeled $\alpha$ and $\beta$. 
In a cylindrical coordinate frame centered on the QD, the electron--hole wave function for the $\alpha$--th QD can be written as
\begin{eqnarray}\label{WF} 
\Psi_\alpha\left({\bf r},{\bf r}\right)&=& \Phi_{\alpha,e}(\phi,\rho,z)\Phi_{\alpha,h}(\phi,\rho,z)\\
&=&\left[e^{im_{\alpha,e}\phi}f_{\alpha,e}(\rho)h_{\alpha,e}(z)\right]\cdot\left[e^{im_{\alpha,h}\phi}f_{\alpha,h}(\rho)h_{\alpha,h}(z)\right]\,.\nonumber
\end{eqnarray}
The details of the calculation of the electron and hole wave functions are given in Appendix~A.

The Maxwell equation for the electric field $\bm{\mathcal E}$, 
expressed in the space and frequency domain, can be written as
\begin{eqnarray} \label{Max}
&&\bm{\nabla}\wedge\bm{\nabla}\wedge
{\bm{\mathcal E}}\left({\bm{\rho}},z,\omega\right)-\frac{\omega^2}{c^2}
\left[\epsilon_\infty {\bm{\mathcal E}}\left({\bm{\rho}},z,\omega\right)\right.
\\
&&\left.+4\pi\int d{\bm{\rho}}^\prime dz^\prime
\hat{\bm{\chi}}\left({\bm{\rho}},{\bm{\rho}}^\prime,z,z^\prime,\omega\right)
\cdot {\bm{\mathcal E}}\left({\bm{\rho}}^\prime,z^\prime,\omega\right)\right]=0
\,,\nonumber
\end{eqnarray} 
where we distinguish between the {\em z}-- and the in--plane ${\bm{\rho}}$ directions. In what follows we omit the $\omega$--dependence in the notation for the electric field, unless required. In Eq.~(\ref{Max}) we assumed a uniform dielectric background with dielectric constant $\epsilon_\infty$, which models the semiconductor matrix surrounding the QD. The in--plane and {\em z}--components of the electric field are defined as ${\bm{\mathcal E}}=\left({\bf E},E_z\right)$. Since $E_z$ is not coupled to the polarization field, it can be easily eliminated from Eq.~(\ref{Max}). The Fourier transform to reciprocal in--plane space is defined as ${{\bf E}\left({\bm{\rho}},z\right)=\sum_{\bf k} {\bf E}_{\bf k}\left(z\right)\mbox{exp}\left[i{\bf k}\cdot{\bm{\rho}}\right]}$. After some algebra, the resulting equation for the in--plane component ${\bf E}_{\bf k}(z)$ reads
\begin{eqnarray} \label{Max_2d}
-\left(1+\frac{1}{k_z^2}\frac{\partial^2}{\partial z^2}\right)
\left(
\begin{array}{cc}
k_0^2-k_y^2   &    k_xk_y \\
k_xk_y        &    k_0^2-k_x^2  
\end{array}
\right)
{\bf E_{\bf k}}\left(z\right)=               \\
4\pi\frac{k_0^2}{\epsilon_\infty}
\sum_{\bf {k^\prime}}\int dz^\prime 
\hat{\bm{\chi}}_{\bf {k,k^\prime}}\left(z,z^\prime\right)\cdot
 {\bf E}_{\bf {k^\prime}}\left(z^\prime\right)\,,\nonumber
\end{eqnarray}
{\bf where}
\begin{eqnarray}
k_z &=& \sqrt{k^2_0-k^2}\label{kz}\\ 
k_0&=&\left(\omega/c\right)\sqrt{\epsilon_\infty}\label{k0}
\end{eqnarray}
are the {\em z}--component of the photon wave vector and the photon dispersion respectively. In what follows, the $\omega$--dependence of the various quantities in the equations is implicitely contained in their $k$--dependence through Eqs. (\ref{kz}) and (\ref{k0}). In Eq.~(\ref{Max_2d}) the susceptibility $\hat{\bm{\chi}}_{\bf {k,k^\prime}}\left(z,z^\prime\right)$ is now a rank--2 tensor acting on the $\left(k_x,k_y\right)$--plane, obtained by Fourier transforming to {\bf k}--space the  $\left(x,y\right)$--minor of the tensor (\ref{Chi}). Eq.~(\ref{Max_2d}) can be solved using the scattering approach proposed in Ref.~\cite{Martin_1998}. The background Green's function of the system is defined as the solution of the left--hand side of Eq.~(\ref{Max_2d}) with an inhomogeneous term $\hat{\bm{I}}\delta\left(z\right)$ on the right--hand side and with outgoing boundary conditions. This Green's function can be derived analytically and reads
\begin{equation} \label{Green}
\hat{\bm{G}}_{\bf k}\left(z\right)=\frac{i}{2k_0^2k_z}
\left(
\begin{array}{cc}
k_0^2-k_x^2   &    -k_xk_y \\
-k_xk_y       &    k_0^2-k_y^2  
\end{array}
\right) {\mbox{exp}}\left[ik_z\left|z\right|\right]\,.
\end{equation}
As already mentioned above, the basis of this two by two tensor corresponds to the $x$ and $y$ directions of the electric field polarization and of the interband optical polarization. The nondiagonal terms have their physical origin in the long--range part of the electron--hole exchange interaction, which is contained in a full Maxwell--Schr\"odinger formalism \cite{Andreani_Review}. For a single QD having cylindrical symmetry, the nondiagonal terms average to zero when evaluating the optical transition amplitude, as expected in an isotropic system. If the system displays an anisotropy, as is the case for two or more QDs, these nondiagonal terms are responsible for the {\em longitudinal--transverse} or fine structure splitting of the resulting polariton modes. 
The Green's function (\ref{Green}) allows to express Eq.~(\ref{Max_2d}) in terms of a Dyson equation as follows
\begin{eqnarray} \label{Dyson}
\lefteqn{{\bf {E_k}}\left(z\right)={\bf E}^0_{\bf k}\left(z\right) +4\pi\frac{k_0^2}{\epsilon_\infty}}\\
&&\times\sum_{\bf {k^\prime}}
\int dz^\prime dz^{\prime\prime}
\hat{\bm{G}}_{\bf k}\left(z-z^\prime\right)\cdot
\hat{\bm{\chi}}_{\bf {k,k^\prime}}\left(z^\prime,z^{\prime\prime}
\right)\cdot {\bf E}_{\bf {k^\prime}}\left(z^{\prime\prime}\right)
\,,\nonumber
\end{eqnarray}
where ${\bf E}^0_{\bf k}$ is the solution of the free propagating field in the dielectric background, namely in the absence of the resonant non--local susceptibility. As already pointed out, we consider cylindrical QDs whose thickness in the $z$--direction is very small compared to their size in the  $\left(x,y\right)$--plane. In this case we can approximate the $z$--dependence of the electron--hole pairs wave functions $\Psi_\alpha$ with a Dirac--delta function. This allows us to rewrite Eq.~(\ref{Dyson}) in the simpler form
\begin{equation} \label{algebric_Dyson}
{\bf {E_k}}={\bf E}^0_{\bf k}+4\pi\frac{k_0^2}{\epsilon_\infty}
\frac{\mu_{cv}^2}{\hbar}\sum_{{\bf {k^\prime}},\beta}
\frac{\psi^{ }_{\beta,\bf{k}}\psi^*_{\beta,\bf{k^\prime}}}
{\omega_\beta-\omega-i0^+}\hat{\bf G}_{\bf k}\cdot 
\bf{E_{\bf {k^\prime}}}\,,
\end{equation}
where all the quantities are defined at the  $\left(x,y\right)$--plane position $z=0$. Here, $\psi_{\beta,{\bf k}}$ is the two--dimensional Fourier transform of $\psi_\beta({\bm{\rho}})=e^{i(m_{\beta,e}+m_{\beta_,h})\phi}f_{\beta,e}(\rho)f_{\beta,h}(\rho)$, that is the in--plane projection of the electron--hole pair wave function in the $\beta$--th QD. If $\bf{R_\beta}$ is the position of the QD in the chosen coordinate frame, then
\begin{equation}
\psi_\beta({\bm{\rho}})=\varphi_\beta({\bm{\rho}}-{\bf{R_\beta}})\,,
\end{equation}
where $\varphi_\beta({\bm{\rho}})$ is the $\beta$--th QD wave function centered at the origin of the coordinate frame. The Fourier transform in k--space then reads
\begin{equation} \label{psi_bK}
\psi_{\beta,\bf{k}}=\varphi_{\beta,\bf{k}}\exp\left[i\bf{k}\cdot\bf{R_\beta}\right]\,.
\end{equation}
Here, because of the cylindrical simmetry of the wave function $\varphi_\beta({\bm{\rho}})$,
\begin{eqnarray} \label{psi_K}
\varphi_{\beta,\bf{k}}&=&\frac{1}{2\pi}\int d{\bm{\rho}}\varphi_\beta({\bm{\rho}})\exp\left(i{\bf{k}}\cdot{\bm{\rho}}\right)\\
&=&\int_0^\infty d\rho \,\rho \varphi_\beta(\rho) J_0(k\rho)\,.\nonumber
\end{eqnarray}
We now project Eq. (\ref{algebric_Dyson}) onto the set of pair wave functions $\psi_{\alpha,{\bf k}}$. The result is
\begin{equation} \label{projected_Dyson}
{\bf E}_\alpha={\bf E}^0_\alpha+\sum_\beta
\frac{\hat{\bf G}_{\alpha\beta}}
{\omega_\beta-\omega-i0^+}{\bf E}_\beta\,,
\end{equation}
where
\begin{equation} 
{\bf E}_\alpha=\sum_{\bf k}{\psi^{ }_{\alpha,\bf{k}}}{\bf E}_k\,,
\end{equation}
\begin{equation}\label{G_ab}
\hat{\bf G}_{\alpha\beta}=4\pi\frac{k_0^2}{\epsilon_\infty}
\frac{\mu_{cv}^2}{\hbar}\sum_{\bf k}{\psi^{ }_{\alpha,\bf{k}}}
\hat{\bf G}_{\bf k}{\psi^{*}_{\beta,\bf{k}}}\,.
\end{equation}
Here, as above, the $\omega$--dependence enters these expressions through the definitions of $k_0$, $k_z$, and $\hat{\bf G}_{\bf k}$. The QD coupling matrix  $\hat{\bf G}_{\alpha\beta}$ is explicitely derived in Appendix B. In particular, in Eq. (\ref{G_int}) the in--plane momentum $k$ is integrated over the whole range, including both radiative modes with $k<k_0$ and surface modes with $k>k_0$. These latter modes, which are evanescent in the $z$ direction, span the largest part of the exchanged photons phase space and are thus ultimately responsible for the transfer mechanism we are describing.
The set of functions $\psi_{\alpha,{\bf k}}$ is in general a non--complete set and therefore, by making this projection, we lose information on the value assumed by the electric field ${\bf E}_{\bf k}$ in all ${\bf k}$--space. Formally, once the quantities ${\bf E}_\alpha$ have been computed, the electric field in all ${\bf k}$--space could in principle be reconstructed by solving again Maxwell equations, using the values ${\bf E}_\alpha$ at each QD as source terms. As it will become clear later, however, the projected values of the electric field are sufficient for the purpose of the present analysis, which is to compute the polariton resonances of the system. It clearly emerges from the structure of Eq. (\ref{projected_Dyson}) that in the absence of coupling, the input field is scattered by each QD individually. Radiative coupling is responsible for the reabsorption of the scattered photons by other QDs, through the terms ${\hat{\bf G}_{\alpha\beta}}$ with $\alpha\neq\beta$. By neglecting these nondiagonal terms we obtain a Dyson equation for a single QD
\begin{equation} \label{diagonal_Dyson}
{\bf E}_\alpha={\bf E}^0_\alpha+
\frac{\hat{\bf G}_{\alpha\alpha}}
{\omega_\alpha-\omega-i0^+}{\bf E}_\alpha\,,
\end{equation}
where $\hat{\bf G}_{\alpha\alpha}=\hat{\bf I}G_\alpha$ ($\hat{\bf I}$, being the 2 by 2 unit matrix), and
\begin{equation} 
G_\alpha = i\frac{2\pi^2\mu_{cv}^2}{\hbar\epsilon_\infty}\int_0^\infty dk |\psi^{ }_{\alpha,\bf{k}}|^2\frac{k(2k_0^2-k^2)}{k_z}\,.
\label{diagonal_SE}
\end{equation}
Eq. (\ref{diagonal_Dyson}) can be solved straightforwardly. The quantity $-G_\alpha$ is the radiative self energy of the $\alpha$-th QD, with its real and imaginary parts describing the radiative energy shift and radiative linewidth (inverse lifetime) respectively. As discussed later, this diagonal approximation already implies an inhomogeneous distribution of the $G_\alpha$, due to the size distribution of the QDs.

\begin{figure}[h]
\begin{center}
\includegraphics[width=8cm]{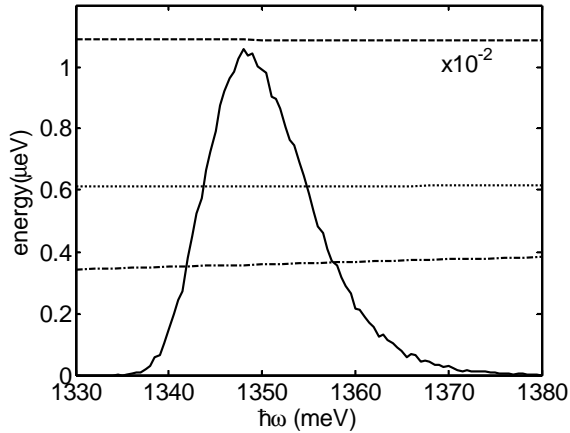}
\caption{QD transition energy distribution (full line, arbitrary units). The asymmetry with a more pronounced high--energy tail is due to $R^{-2}$--dependence of the $\alpha$--th QD confinement energy on the QD radius $R_\alpha$, the radii being Gauss--distributed. The $(x,x)$-component of the coupling energy tensors $\Re\{G_\alpha\}$ (dashed), $\Re\{G^{xx}_{\alpha\beta}\}$ (dotted), and $\Im\{G_\alpha\}$ (dot-dashed), for two QDs labelled $\alpha$ and $\beta$, is plotted as a function of $\hbar\omega$.}
\label{QDdistr}
\end{center}
\end{figure}

In this work we are interested in the effect of radiative coupling between distant QDs. To this purpose, we seek for the solutions of the coupled Dyson equation (\ref{projected_Dyson}). The polariton resonances of the multiple--QD system are then the poles of the homogeneous set of equations obtained by setting ${\bf E}^0_\alpha=0$ in Eqs. (\ref{projected_Dyson}). We compute these poles numerically within the exciton--pole approximation \cite{Tassone_NCimento_1990,Citrin_PRB_1993,Jorda_PRB_1993,Jorda_PRB_1994}, which consists in replacing the $\omega$--dependence of ${\hat{\bf G}_{\alpha\beta}}$ tensor by an average electron hole energy $\hbar\omega_0$. This approximation is generally valid when the dielectric medium does not present sharp resonances, as is the case in the present model where the QDs are embedded in a constant dielectric background. In order to check the validity of this assumption, we evaluated the $\omega$--dependence of the coupling tensor ${\hat{\bf G}_{\alpha\beta}}$ for a pair of QDs and checked that all its components are essentially constant over the energy interval corresponding to a typical inhomogeneous QD distribution. Some of these components are plotted in Fig. \ref{QDdistr} as a check. Complex eigenenergies $\Omega_n = \Delta_n + i\Gamma_n$ are obtained, corresponding to collective radiative modes of the QD ensemble. The number of these poles is twice the number of QDs, corresponding to the two independent states of the interband polarization vector. The real part  of the n--th eigenvalue $\Delta_n$ induces a radiative shift with respect to the energies of the non--interacting dots, while the imaginary part $\Gamma_n$ represents the radiative recombination rate of the n--th collective mode of the system.
\section{Numerical results}

In the first part of this section we will address a two--QD system, in order to establish how the radiative coupling mechanism depends on the {\em{detuning}} and on the mutual QD distance. Here, the {\em{detuning}} is defined as the difference between the optical transition energies of the two QDs. In the second part, we will discuss the results obtained for an ensemble of several dots. In order to have a quantitave estimate of the effect, we will show results relative to the realistic cases of an InAs QD ensemble \cite{Stier_PRB_1999,Grundmann_PRB_1995} and of a CdSe one \cite{Litvinov02}, which differ from each other for the values of the dipole matrix element $\mu_{cv}$ and for the QDs spatial density in typical samples.  

\begin{figure}[h]
\begin{center}
\includegraphics[width=8cm]{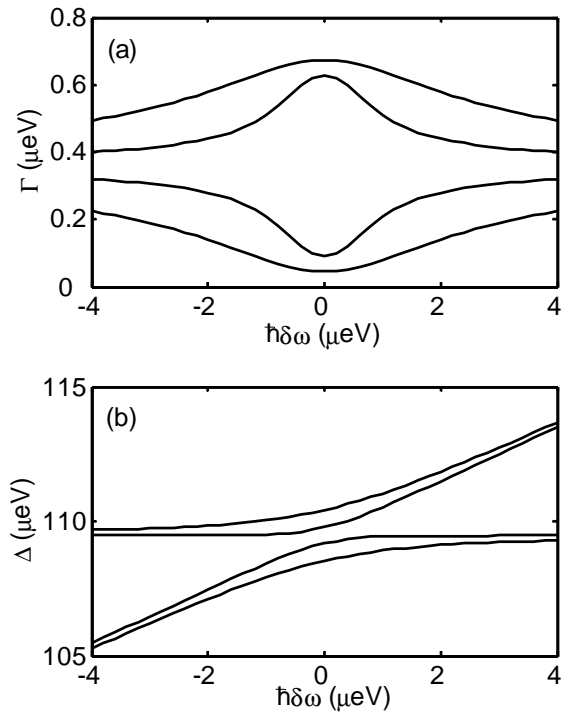}
\caption{Imaginary~(a) and real~(b) part of the energy poles as a function of the detuning between two QDs, at fixed distance $R_{\alpha\beta}=50$~nm.  The energy scale is relative to the case of InAs QDs with $\mu_{cv}^2=480\mbox{ meV/nm}^3$ and a radius of the cylinder of about 10 nm. Note that for small detuning the four poles are well separed in energy, so that in (a) two sub--radiant and two super--radiant states are distinguishable.} 
\label{2QD_vs_omega}
\end{center}
\end{figure}

\begin{figure}[h]
\begin{center}
\includegraphics[width=8cm]{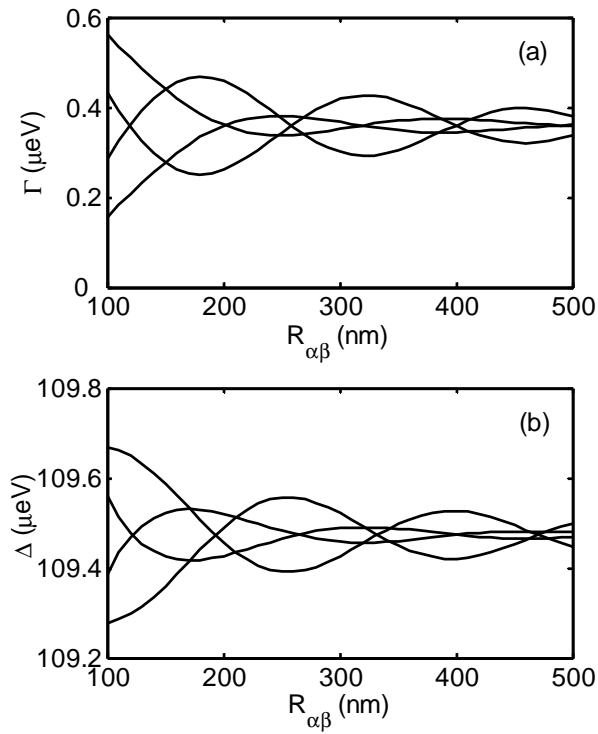}
\caption{Imaginary~(a) and real~(b) part of the energy poles as a function of the distance between two QDs, at zero detuning.  The energy scale is the same as in  Fig.~\ref{2QD_vs_omega}. The oscillatory nature of the interaction as a function of distance, according to the Bragg or anti--Bragg condition, clearly appears.}
\label{2QD_vs_D}
\end{center}
\end{figure}

We first consider the case of two QDs. The QDs are assumed of cylindrical shape. The electron and hole wave functions are calculated within the effective mass approximation, assuming a finite barrier at the QD boundaries (see Appendix A). The cylindrical shape enables us to analitically derive the elements of the $\hat{\bf G}_{\alpha\beta}$ tensor in Eq.~(\ref{projected_Dyson}) and simplifies our numerical task. The detuning is changed by varying the size of one of the QDs. In Fig.~\ref{2QD_vs_omega} the imaginary~(a) and the real~(b) part of the poles of Eq.~(\ref{projected_Dyson}) (that is,  $\Gamma_n$ and  $\Delta_n$ respectively) are plotted versus the detuning $\hbar\delta\omega=\hbar(\omega_1-\omega_2)$ of the two QDs, at fixed distance.  The energy scale is relative to the physical parameters of Stranski--Krastanov grown InAs QDs, that is a dipole matrix element $\mu_{cv}^2=480\mbox{ meV/nm}^3$, corresponding to a Kane energy of 22~eV~\cite{Vurgaftman}, and a radius of the cylinder of about 10 nm. The numerical simulations show that no appreciable coupling effect is observed for large detuning, as expected. On the other hand, for small detuning the energies of the four poles are well distinguished. In particular, if we look at $\Gamma_n$ in Fig.~\ref{2QD_vs_omega}(a), we can see that two {\em{sub--radiant}} and two {\em{super--radiant}} states are present. The two states with small $\Gamma_n$, thus, decay in a time much longer than the two others. The computed energy shift with respect to non--interacting dots is of the same order of $\Gamma_n$, that is of the order of 1~$\mu$eV. Such an energy shift is negligible if compared to the typical inhomogeneous broadening of a QD ensemble. The main consequence of radiative coupling is thus the effect on the lifetimes of the collective modes of the system.
Fig.~\ref{2QD_vs_D} displays the dependence of the interaction on the distance between the QDs. The imaginary~(a) and the real~(b) part of the poles oscillate as a function of the distance between the two dots. The oscillations originate from interference effects. At distances which are multiple of the half wavelength, Bragg or anti--Bragg conditions are satisfied and the oscillations display a maximum or a node, respectively. Fig. \ref{2QD_vs_D} illustrates the long-range character of this radiative coupling mechanism. The magnitude of the coupling, expressed as the envelope of the curves in Fig. \ref{2QD_vs_D} (a) and (b), can be inferred from Eq. (\ref{G_int_ang}) and decreases as $(R_{\alpha\beta})^{-1}$, where $R_{\alpha\beta}$ is the distance between the dots. As already pointed out, this dependence is much slower than the characteristic $(R_{\alpha\beta})^{-6}$ dependence of the F\"orster coupling \cite{Foerster,Govorov03}. It should be pointed out that our theory makes use of the Coulomb gauge for the Maxwell equations and in particular for the dipole Hamiltonian from which the linear susceptibility is derived. In this limit, only transverse fields are considered and the electrostatic interaction, which is related to the instantaneous longitudinal part of the electromagnetic field, is excluded from the treatment \cite{Jackson}. In a very recent work \cite{Sangu04}, the same process of energy transfer by emission and reabsorption of a photon has been described in the instantaneous limit, by using second order perturbation theory for the derivation of the transfer rate, without introducing the Maxwell equations. In this limit, the interaction turns out to decay exponentially with the distance, a result which is well expected as the radiative nature of the interaction is neglected.

\begin{figure}[h]
\begin{center}
\includegraphics[width=8cm]{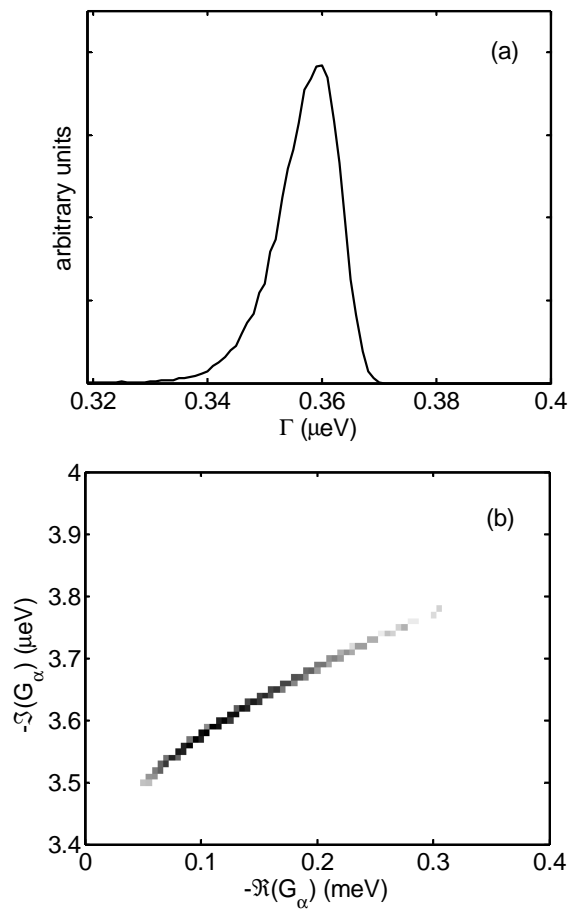}
\caption{(a) Histogram representing the energy distribution of the single QD radiative rates, expressed as the imaginary part of the single QD radiative self--energy $\gamma_\alpha=-\Im\{G_\alpha\}$. (b) Two--dimensional histogram of the distribution of the real and imaginary parts of the single--QD radiative self--energy.}
\label{QDg0}
\end{center}
\end{figure}

\begin{figure}[h]
\begin{center}
\includegraphics[width=8cm]{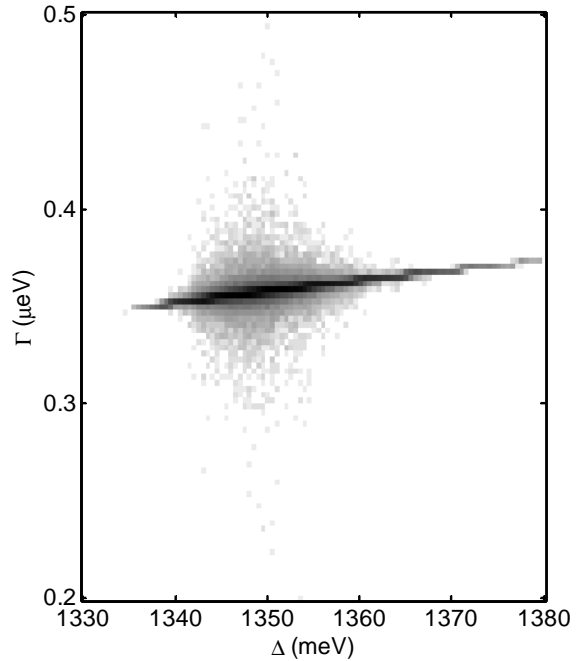}
\caption{Logarithmic scale histogram expressing the number of collective modes as a function of the real and imaginary part of the complex energy poles of an InAs QDs ensemble. The physical parameters of the QDs are the same as in  Fig.~\ref{2QD_vs_omega} and their density is 300~QDs/$\mu\mbox{m}^2$. A fraction of the QDs shows however a large radiative shift.}
\label{InAs}
\end{center}
\end{figure}

\begin{figure}[h]
\begin{center}
\includegraphics[width=8cm]{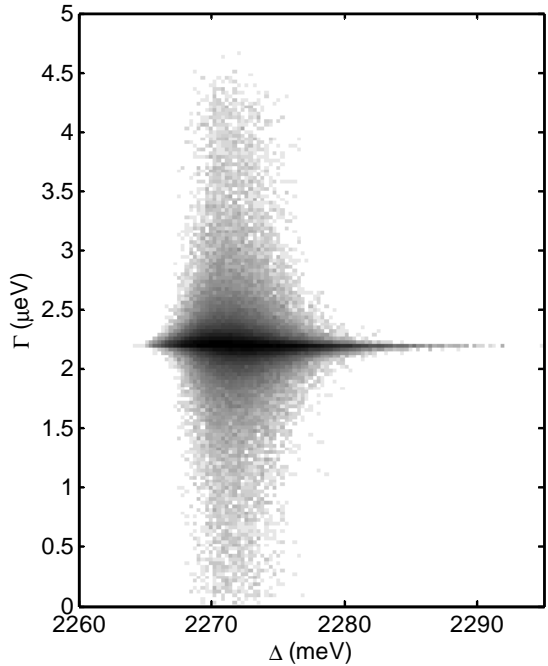}
\caption{The same histogram of Fig.~\ref{InAs}, but for GaAs QDs with $\mu_{cv}^2=780\mbox{ meV/nm}^3$ and a physical density of 1000~QDs/$\mu\mbox{m}^2$. In this case radiative shifts up to one order of magnitude larger than in the case of InAs QDs are obtained as an effect of the coupling.}
\label{CdSe}
\end{center}
\end{figure}

Now that the features of the radiative coupling mechanism have been clarified, we consider the case of a large number of interacting QDs. The same emission--absorption mechanism that couples a pair of dots can now involve several QDs and the transfer of excitation between them results in collective modes analogous to the ones previously described, that is, sub--radiant or super--radiant if compared to the excited states of the 
non--interacting dots. As a first example, we continue to use the parameters of InAs QDs, which are randomly distributed in the $\left(x,y\right)$--plane, with a physical density of 300~QDs/$\mu\mbox{m}^2$. In a real situation, the dots have different shape, size and composition causing the inhomogeneous energy broadening of the QD luminescence spectrum. To simulate this broadening, we introduce a Gauss distributed dot  size centered at dot radius $R=10\mbox{ nm}$, with a standard deviation of $\delta R=1\mbox{ nm}$. This variance in size induces a variation of the confinement energy $\hbar\delta\omega$ which is proportional to $\delta R/R^3$ as implied by the energy quantization of a particle in a box. This energy variation is what finally produces the inhomogeneous energy distribution of the QDs. The choice $\delta R=1\mbox{ nm}$, given our simple model for the QD wave functions, results in an inhomogeneous broadening of about 15 meV, as seen in Fig. \ref{QDdistr}. The asymmetry of this distribution, with a more pronounced high--energy tail, is simply related to the $R^{-3}$--dependence of the confinement energy variation and to the Gauss assumption for the distribution of QD sizes. The same size fluctuation is also responsible of a variation of the QD optical matrix element \cite{Borri_PRB_2002} and consequently of both its radiative shift and lifetime, via Eq. (\ref{diagonal_Dyson}) and the single--dot self--energy (\ref{diagonal_SE}). The numerically computed radiative energy shifts are of the order of a few $\mu$eV, thus negligible if compared to the QD inhomogeneous energy broadening. They are therefore irrelevant to the present discussion. The imaginary part of the single--dot self--energy is on the contrary what gives the inhomogeneous distribution of radiative linewidths $\gamma_\alpha=-\Im\{G_\alpha\}$. Their distribution is plotted in Fig. \ref{QDg0}(a). Finally, Fig. \ref{QDg0}(b) shows a two-dimensional histogram of $-\Re\{G_\alpha\}$, and $-\Im\{G_\alpha\}$, showing the correlation between radiative shift and radiative broadening resulting from the present model. In a realistic situation \cite{Borri_PRB_2002}, a variation of the dipole moment is not only induced by size fluctuations. Other factors such as QD shape, strain and piezoelectric fields, and indium concentration within the QD body produce a variation of dipole moment even for a fixed QD size. The 20\% variance of the dipole moments derived in Ref \cite{Borri_PRB_2002} is significantly larger than the one obtained here from size fluctuations (approx. 3\% for the InAs case). However we note that the inhomogeneous broadening of the sample by Borri et al. is also larger than the one considered here, presumably due to an even larger QD--size fluctuation. Introducing a larger size fluctuation in the present model would partly account for the observed dipole--moment fluctuation. Our final result for a radiatively--coupled QD ensemble however (see discussion below and Figs. \ref{InAs}, and \ref{CdSe}), predicts an even broader distribution of radiative linewidths which might be at least partly responsible for the measured dipole moment distribution.

We compute the collective modes of an ensemble of 100 QDs by finding the complex poles of Eq. (\ref{projected_Dyson}). We repeat this procedure for many random realizations of the system. Provided the system size is larger than the wavelength, we expect this configuration average to give the same results as a simulation over a larger spatial domain. This is true because of the fall--off scale computed in Fig. \ref{2QD_vs_D}. In particular, the occurrence of quasi--degenerate QD pairs within a given realization has a finite though small probability. Repeating the simulation over many randomly generated configurations finally allows to sample over a large enough number of such quasi--degenerate cases and produces a significant statistics. We plot in Fig.~\ref{InAs} an histogram, on a logarithmic scale, of the real and imaginary parts of the computed energy poles. Most of the collective modes lie on the curve determined by the distribution of non--interacting QDs displayed in Fig. \ref{QDg0}(b), due to the large detunings that are induced by the inhomogeneity of the QD ensemble. Nevertheless, for a small fraction of the states a large radiative shift is achieved, as a result of the coupling. We also point out that the deviation from the non--interacting QDs case is more pronunced in correspondence of the center of the QD inhomogeneous line. The reason is that, as already stated, the radii of the QDs are Gauss--distributed around a mean value. Most of the QDs fall in this energy region and consequently small values of the detuning are more likely to occur.

For different materials, one can observe larger radiative coupling effects. In Fig.~\ref{CdSe} we show the histogram obtained for a CdSe QD sample. The physical parameters for this case  are a spatial density of 1000~QDs/$\mu\mbox{m}^2$ and a dipole matrix element value of $\mu_{cv}^2=780\mbox{ meV/nm}^3$. Once again the histogram results from many realization of the sample, with randomly distributed QDs, having ramdomly Gauss--distributed size. In this case the deviation from the noninteracting case is more pronounced because of the higher QD density and of the larger dipole matrix element $\mu_{cv}$. A radiative shift of a few $\mu$eV is achieved, that is one order of magnitude larger with respect to the case of InAs QDs. In this case, some of the collective modes have vanishing radiative rates, showing how the radiative coupling can profoundly change the dephasing rates of many QD systems.
\section{Conclusions}
We have shown that QDs in a sample cannot in principle be considered as isolated 
systems. The radiative coupling between QDs causes the emergence of collective 
modes. By comparing their lifetimes with the ones of the excited state of an 
isolated QD, we can classify these modes into sub--radiant and super--radiant. 
We find the effect on the radiative decay--rate to be of the order of 1~$\mu$eV. 
This effect strongly depends on the dipole matrix element of the material that 
constitutes the QDs and on their spatial density. For a very dense QD sample 
this effect should be observable as a non--exponential decay of the 
photoluminescence signal. Despite its small size, the addressed mechanism acts 
over wavelength distance, so that two QDs that are a few hundreds of nanometers 
far from each other can radiatively interact. Semiconductor QDs are being 
increasingly advertised as the ideal building blocks of the future technology 
for quantum information processing. These proposals are often based on pairs 
of identical QDs \cite{Quiroga99} or on pairs of QDs in which a degeneracy 
occurs between different excited levels \cite{Govorov03} and often take 
advantage of excitation transfer processes. Moreover, it is likely that a solid 
state implementation of a quantum information system would be constituted of a 
great number of (nearly) identical, independent quantum gates, possibly located 
at submicron distance from each other. In all these situations where levels of 
different QDs are nearly degenerate, our result shows that excitation transfer 
by radiative coupling can occur over long distances. The radiative coupling 
mechanism that we describe might therefore be relevant in determining the 
excitation transfer dynamics of these systems.

\begin{acknowledgments}
We are grateful to A. Quattropani, P. Schwendimann, and R. Zimmermann for fruitful discussions. We acknowledge financial support from the Swiss National Foundation through project N. 620-066060.
\end{acknowledgments}
\appendix
\section{Carrier wave functions}

Due to the symmetry of the problem, in the following we will consider a cylindrical coordinate system ($\phi,\rho,z$). The in--plane radius of the cylindrical QD is $R$, and its height in the $z$--direction is $h$. The effective mass Hamiltonian operator which describes the carrier (electron or hole) in the QD is
\begin{equation}\label{hamiltonian}
H=-\frac{\hbar^2{\bm{\nabla}}^2}{2m_c}+V(\rho,z)\,,
\end{equation}
where $m_c$ is the effective mass of the carrier and $V(R,z)$ describes the band profile for the QD, that is
\begin{equation}
V(\rho,z)=\left\{
\begin{array}{rl}
0 & ,\,\,\rho< R \,\,\mbox{and} \,\,|z|<h/2\\
V & ,\,\,\mbox{elsewhere}
\end{array}\,.
\right.
\end{equation}

Three assumptions allow to simplify the problem:
(i) we assume that the problem is separable, namely the wave--function can be written as 
\begin{equation}\label{wave_function}
\Phi(\phi,\rho,z)=e^{im\phi}f(\rho)h(z)\,, 
\end{equation}
where $m$ is a positive integer representing the azimuthal quantum number;
(ii) we assume that the effective mass of the carrier is the same in the QD and in the surrounding medium; 
(iii) we rewrite the Hamiltonian operator as
\begin{eqnarray}
H &=& H_0+H_1\,,     \\
H_0 &=& -\frac{\hbar^2{\bm{\nabla}}^2}{2m_c} + U(\rho) +W(z)\,,  \\
H_1 &=& V(\rho,z) - U(\rho) - W(z)\,.
\end{eqnarray}
with
\begin{eqnarray}
%\begin{array}{rl}
U(\rho)=\left\{
\begin{array}{ll}
0 & , \,\,\rho\leq R \\
V &  , \,\,\rho> R 
\end{array}
\right.   \,,            \\
W(z)=\left\{
\begin{array}{ll}
0 & ,\,\,|z|\leq h/2\\
V & ,\,\,|z|> h/2
\end{array}
\right.\,.
%\end{array}
\end{eqnarray}
$H_1$ is considered as a small perturbation, because it is nonzero only in  regions  of space where the $R$-- and $z$--confined wave--functions assume very small values.

By neglecting $H_1$, the problem becomes separable. In the $z$--direction we reduce to the problem of the one--dimensional square potential. Because of the symmetry of the problem, we find both even and odd solutions.

The even solution is
\begin{equation}
h(z)=\left\{
\begin{array}{ll}
N\cos(kz) & , \,\,|z|< h/2\\
N\cos(kh/2)e^{-k^\prime(z-h/2)} & , \,\,z> h/2\\
N\cos(kh/2)e^{k^\prime(z+h/2)} & , \,\,z< -h/2
\end{array}
\right.\,,
\end{equation}
where $k^\prime=\sqrt{2m_cV/\hbar^2-k^2}$ and $N$ is a normalization factor.
The conditions of continuity of the solution and of its derivative  
require that $k$ verifies the equation
\begin{equation}\label{cont_1}
k\tan{\left(\frac{kh}{2}\right)}=k^\prime\,.
\end{equation}

The odd solution is
\begin{equation}
h(z)=\left\{
\begin{array}{ll}
N\sin(kz) & , \,\,|z|< h/2\\
N\sin(kh/2)e^{-k^\prime(z-h/2)} & , \,\,z> h/2\\
-N\sin(kh/2)e^{k^\prime(z+h/2)} & , \,\,z< -h/2
\end{array}
\right.\,.
\end{equation}
In this case imposing the continuity at the QD boundaries implies
\begin{equation}\label{cont_2}
k\cot{\left(\frac{kh}{2}\right)}=-k^\prime\,.
\end{equation}
Eqs.~\ref{cont_1} and \ref{cont_2} result in a discretization of the wave vector, which will be labeled by $n$.

The radial part of the Scr\"{o}dinger equation takes the form of a Bessel equation
\begin{equation}\label{radial_Schroedinger}
f^{\prime\prime}(R)+\frac{1}{R}f^\prime(R)+\left(A-\frac{2m_c}{\hbar^2}U(R)-\frac{m_c^2}{R^2}\right)f(R)=0\,.
\end{equation}
Solutions of this equation are the Bessel functions. In the QD the wave function must be well defined at $R=0$, while outside of the QD we look for exponentially decaying solutions, as required for a confined state. These requirements are satisfied by first kind Bessel functions and first Hankel functions with imaginary argument, respectively. We obtain
\begin{equation}
f(\rho)=\left\{
\begin{array}{ll}
N^\prime J_m(q\rho) & ,\,\,\rho<R\\
N^\prime J_m(q\rho)H_m^1(q^\prime\rho)/H_m^1(q^\prime R) & , \,\,\rho>R\\
\end{array}
\right.\,,
\end{equation}
where $q^\prime=i\sqrt{2m_cV/\hbar^2-q^2}$, $q^2<2m_cV/\hbar^2$ and $N^\prime$ is a normalization factor. The conditions of continuity are achieved if $q$ satisfies the equation
\begin{equation}
q\frac{J_{m-1}(qR)-J_{m+1}(qR)}{J_{m}(qR)}=q^\prime\frac{H_{m-1}^1(q^\prime R)-H_{m+1}^1(q^\prime R)}{H_{m}^1(q^\prime R)}\,,
\end{equation}
resulting in the discretization of $q$, which we label by $l$. The problem has therefore three quantum numbers, namely $l$, $m$ and $n$.

We have evaluated, at the first order of perturbation, the error introduced by neglecting $H_1$. This error is less than 1\% 
for the confined functions, that is negligible also considering the other approximations made.

The excitonic wave--function is the product of the ground--state wave--functions of electron and hole, namely the functions corresponding to $l=1$, $m=0$ and $n=1$. A first improvement of the model, aimed at taking into account the Coulomb interaction, would consist in a variational approach based on a linear superposition of $(l,m,n)$--states  with the coefficients chosen to minimize the Coulomb interaction.

\section{QD coupling tensor}
Using the expression (\ref{psi_bK}) for the electron--hole pair wave function in $k$--space, the coupling tensor $\hat{\bf G}_{\alpha\beta}$ in Eq.~(\ref{G_ab}) becomes
\begin{equation}\label{G_sum}
\hat{\bf G}_{\alpha\beta}=4\pi\frac{k_0^2}{\epsilon_\infty}
\frac{\mu_{cv}^2}{\hbar}\sum_{\bf k}{\varphi^{ }_{\alpha,\bf{k}}}{\varphi^{*}_{\beta,\bf{k}}}
\hat{\bf G}_{\bf k}{\exp\left[-i\bf{k}\cdot\bf{R_{\alpha\beta}}\right]}\,,
\end{equation}
where $\bf{R_{\alpha\beta}}=\bf{R_\alpha}-\bf{R_\beta}$ is the distance vector between QDs $\alpha$ and $\beta$. Turning the sum into an integral, Eq.~(\ref{G_sum}) can be written as
\begin{eqnarray}\label{G_int}
\hat{\bf G}_{\alpha\beta}&=&i\frac{2\pi}{\hbar}\frac{\mu_{cv}^2}{\epsilon_\infty}
\int_0^\infty dk \frac{k}{k_z}{\varphi^{ }_{\alpha,\bf{k}}}{\varphi^{*}_{\beta,\bf{k}}}\\
&\times&\int_0^{2\pi}d\phi
\left(
\begin{array}{cc}
k_0^2-k^2\cos^2\phi & -k^2\sin\phi\cos\phi \\
 -k^2\sin\phi\cos\phi &  k_0^2-k^2\sin^2\phi 
\end{array}
\right)\nonumber\\
&\times&\exp\left[-ikR_{\alpha\beta}\cos(\phi-\theta_{\alpha\beta})\right]
\,,
\end{eqnarray}
where $\phi$ and $\theta_{\alpha\beta}$ are the angles that the vectors $\bf{k}$ and $\bf{R_{\alpha\beta}}$ respectively form with the $x$--axis of the chosen coordinate frame. For each QD pair $(\alpha,\beta)$ we perform a rotation of the $2\times 2$ matrix in Eq.~(\ref{G_int}) by an angle $\theta_{\alpha\beta}$. The rotation matrix is $\hat{\bf R}_{\theta_{\alpha\beta}}$. In the new coordinate frame the two QDs lie on the $x$--axis. In the rotated frame the expression for the new coupling tensor $\hat{\bf G^{\prime}}_{\alpha\beta}=\hat{\bf R}_{\theta_{\alpha\beta}}\hat{\bf G}_{\alpha\beta}\hat{\bf R}^{-1}_{\theta_{\alpha\beta}}$ is identical to Eq.~(\ref{G_int}), with $\phi-\theta_{\alpha\beta}$ replacing $\phi$ everywhere except in the argument of the exponential. The angular integration can be performed analytically and results in a diagonal matrix as expected 
\begin{widetext}
\begin{eqnarray}\label{G_int_ang}
\hat{\bf G^{\prime}}_{\alpha\beta}&=&
\left(
\begin{array}{cc}
 g^{L}_{\alpha\beta} & 0 \\
0 & g^{T}_{\alpha\beta} 
\end{array}
\right)
=i2\pi\frac{\mu_{cv}^2}{\hbar\epsilon_\infty}\int_0^\infty dk \frac{k}{k_z}{\varphi^{ }_{\alpha,\bf{k}}}{\varphi^{*}_{\beta,\bf{k}}}\\
&& \times
\left(
\begin{array}{cc}
2\pi k_z^2 J_0(kR_{\alpha\beta})+\frac{4\sqrt{\pi}}{R_{\alpha\beta}}\Gamma\left(\frac{3}{2}\right)kJ_1(kR_{\alpha\beta}) & 0 \\
0 &2\pi k_0^2 J_0(kR_{\alpha\beta})-\frac{4\sqrt{\pi}}{R_{\alpha\beta}}\Gamma\left(\frac{3}{2}\right)kJ_1(kR_{\alpha\beta})   
\end{array}
\right)
\,,\nonumber
\end{eqnarray}
\end{widetext}
where $J_n(x)$ is the $n$--th order Bessel function of the first kind and 
$\Gamma(x)$ is the Euler gamma function. Labels ``L'', ``T'' denote the longitudinal and transverse polarizations with respect to the $\bf{R_{\alpha\beta}}$ axis. The expression in Eq.~(\ref{G_int_ang}) 
depends only on the distance between the pair of QDs considered. The integral 
over $k$ is performed numerically. The result is then rotated back by an angle 
$-\theta_{\alpha\beta}$ to obtain the complete coupling matrix in the original 
coordinate frame. The coupling tensor between QDs $\alpha$ and $\beta$ then 
reads
\begin{widetext}
\begin{eqnarray}
\hat{\bf G}_{\alpha\beta}&=&
\left(
\begin{array}{cc}
g^{L}_{\alpha\beta}\cos^2(\theta_{\alpha\beta}) +  g^{T}_{\alpha\beta}\sin^2(\theta_{\alpha\beta}) &  (g^{L}_{\alpha\beta}-g^{T}_{\alpha\beta}) \sin(\theta_{\alpha\beta})\cos(\theta_{\alpha\beta})\\
 (g^{L}_{\alpha\beta}-g^{T}_{\alpha\beta}) \sin(\theta_{\alpha\beta})\cos(\theta_{\alpha\beta}) &  g^{L}_{\alpha\beta}\sin^2(\theta_{\alpha\beta}) +  g^{T}_{\alpha\beta}\cos^2(\theta_{\alpha\beta})
\end{array}
\right)\,.\label{Gab}
\end{eqnarray}
\end{widetext}

\bibliographystyle{PRSTY}

\end{document}